\documentclass[twocolumn,aps,prc,showpacs,floatfix]{revtex4}
\usepackage{graphicx}
\usepackage{dcolumn}
\usepackage{bm}

\begin{document}

\title{Can the nuclear symmetry potential at supra-saturation densities be negative?}
\author{Gao-Chan Yong}
\affiliation{Institute of Modern Physics, Chinese Academy of
Sciences, Lanzhou 730000, China}

\begin{abstract}
In the framework of an Isospin-dependent
Boltzmann-Uehling-Uhlenbeck (IBUU) transport model, for the
central $^{197}$Au+$^{197}$Au reaction at an incident beam energy
of $400$ MeV/nucleon, effect of nuclear symmetry potential at
supra-saturation densities on the pre-equilibrium clusters
emission is studied. It is found that for the positive symmetry
potential at supra-saturation densities the neutron to proton
ratio of lighter clusters with mass number $A\leq3$
($(n/p)_{A\leq3}$) is larger than that of the weighter clusters
with mass number $A>3$ ($(n/p)_{A>3}$), whereas for the negative
symmetry potential at supra-saturation densities the
$(n/p)_{A\leq3}$ is \emph{smaller} than that of the $(n/p)_{A>3}$.
This may be considered as a probe of the negative symmetry
potential at supra-saturation densities.

\end{abstract}

 \maketitle


Recently the studies of the density-dependent nuclear symmetry
energy, which is crucial for understanding many interesting issues
in both nuclear physics and astrophysics
\cite{Bro00,Bar05,LCK08,Sum94,Lat04,Ste05a,piek08,horow01}, have
attracted much attention \cite{xiao09,yong06,LiBA02}. The high
density behavior of the symmetry energy, however, has been
regarded as the most uncertain property of dense neutron-rich
nuclear matter \cite{Kut94,Kub99}. Many microscopic and/or
phenomenological many-body theories using various interactions
predict that the symmetry energy increases continuously at all
densities. On the other hand, other models predict that the
symmetry energy first increases to a maximum and then may start
decreasing at certain supra-saturation densities. Thus, currently
the theoretical predictions on the symmetry energy at
supra-saturation densities are extremely diverse. To make further
progress in determining the symmetry energy at supra-saturation
densities, what are most critically needed is some guidance from
dialogues between experiments and the predictions of heavy-ion
collisions transport models, which have been done extensively in
the studies of nuclear symmetry energy at low densities
\cite{tsang09,shetty07,fami06,tsang04,chen05}.

While studying the symmetry energy by using heavy-ion collisions,
the related input in an IBUU transport model is actually the
symmetry potential, which is a more complete information than the
density dependence of the symmetry energy at zero temperature
calculated with the same (mean-field) approximation.
Unfortunately, the symmetry potential is also rather uncertain,
which can be positive or negative at supra-saturation densities
\cite{liba03,fuchs,zuo,Bar05,beh05,liqf05,liz06}. To study the
symmetry energy at supra-saturation densities in the framework of
an IBUU model, one has to firstly determine the symmetry potential
at higher densities. The symmetry potential, its positive or
negative, thus urgently needs to be solved. In this paper, we use
the asymmetry of cluster emission in dynamical simulation to study
the symmetry potential at supra-saturation densities.

The non-equal partition of the system's isospin asymmetry with the
gas phase being more neutron-rich than the liquid phase has been
found as a general phenomenon using essentially all
thermodynamical models and in simulations of heavy-ion reactions
\cite{das,Bar05,chomaz,li00,xu}. But all these studies are for low
energy density nuclear matter, in which the sub-saturation
symmetry energy/potential dominates the isospin fractionation. For
supra-saturation symmetry energy/potential's studies, one needs to
study the isospin fractionation of higher energy density nuclear
matter. Such matter is explored in the first step of high energy
heavy-ion collisions, where time dependence and out-of-equilibrium
effects play a dominant role. Therefore, to study the high density
behavior of nuclear symmetry energy/potential, one needs to base
on the transport model, to study particle emission by using
relative high incident beam energy of heavy-ion collisions and
compare with the experimental data. In the present study, using an
isospin and momentum-dependent transport model IBUU, as an
example, we studied the neutron-rich reaction of
$^{197}$Au+$^{197}$Au at a beam energy of $400$ MeV/nucleon with
the positive and negative symmetry potentials at supra-saturation
densities while keeping the low density symmetry energy/potential
fixed. We compute the average isospin ratio of clusters of size
larger (smaller) than A=4 (corresponding mass number of $\alpha$
particle), which will be called $(n/p)_{A>3}$ ($(n/p)_{A\leq3}$)
in the following. We find that for the positive symmetry potential
at supra-saturation densities the $(n/p)_{A\leq3}$ is larger than
that of the $(n/p)_{A>3}$, whereas for the negative symmetry
potential at supra-saturation densities the $(n/p)_{A\leq3}$ is
smaller than that of the $(n/p)_{A>3}$.

\begin{figure}[th]
\begin{center}
\includegraphics[width=0.4\textwidth]{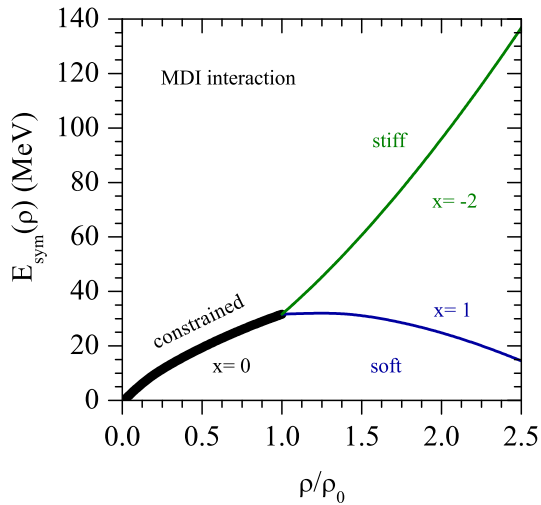}
\end{center}
\caption{The symmetry energy as a function of density in the MDI
interaction with different $x$ parameters. The symmetry energy at
sub-saturation densities was roughly constrained from recent
studies \cite{shetty07,piek08,lit07,todd05}.} \label{esym}
\end{figure}

The isospin and momentum-dependent mean field potential (MDI) used
in the present work is \cite{Das03}
\begin{eqnarray}\label{mdi}
U(\rho ,\delta ,\mathbf{p},\tau )&=&A_{u}(x)\frac{\rho _{\tau ^{\prime }}}{%
\rho _{0}}+A_{l}(x)\frac{\rho _{\tau }}{\rho _{0}}
+B(\frac{\rho }{\rho _{0}})^{\sigma }(1-x\delta ^{2})\nonumber \\
&&-8x\tau \frac{B}{%
\sigma +1}\frac{\rho ^{\sigma -1}}{\rho _{0}^{\sigma }}\delta \rho
_{\tau^{\prime }}  \nonumber \\
&&+\frac{2C_{\tau ,\tau }}{\rho _{0}}\int d^{3}\mathbf{p}^{\prime }\frac{%
f_{\tau}(\mathbf{r},\mathbf{p}^{\prime})}{1+(\mathbf{p}-\mathbf{p}^{\prime
})^{2}/\Lambda ^{2}}\nonumber \\
&&+\frac{2C_{\tau ,\tau ^{\prime }}}{\rho _{0}}\int d^{3}\mathbf{p}^{\prime }%
\frac{f_{\tau ^{\prime }}(\mathbf{r},\mathbf{p}^{\prime })}{1+(\mathbf{p}-%
\mathbf{p}^{\prime })^{2}/\Lambda ^{2}},
\end{eqnarray}
where $\delta =(\rho _{n}-\rho _{p})/\rho$ is the isospin
asymmetry of the nuclear medium. In the above $\tau =1/2$ ($-1/2$)
for neutrons (protons) and $\tau \neq \tau ^{\prime }$; $\sigma
=4/3$; $f_{\tau }(\mathbf{r},\mathbf{p})$ is the phase space
distribution function at coordinate $\mathbf{r}$ and momentum
$\mathbf{p}$. The parameters $A_{u}(x),A_{l}(x),B,C_{\tau ,\tau
},C_{\tau ,\tau ^{\prime }}$ and $\Lambda $ were obtained by
fitting the momentum-dependence of the $U(\rho ,\delta
,\mathbf{p},\tau ,x)$ to that predicted by the Gogny Hartree-Fock
and/or the Brueckner-Hartree-Fock (BHF) calculations \cite%
{bombaci}, the saturation properties of symmetric nuclear matter
and the symmetry energy of about $30$ MeV at normal nuclear matter
density $\rho _{0}=0.16$ fm$^{-3}$ \cite{Das03,liba03}. The
incompressibility $K_{0}$ of symmetric nuclear matter at $\rho
_{0}$ is set to be $211$ MeV consistent
with the latest conclusion from studying giant resonances \cite%
{k0data,pie04,colo04}. The parameters $A_{u}(x)$ and $A_{l}(x)$
depend on the $x$ parameter according to
\begin{equation}
A_{u}(x)=-95.98-x\frac{2B}{\sigma
+1},~A_{l}(x)=-120.57+x\frac{2B}{\sigma +1}.
\end{equation}
The variable $x$ is introduced to mimic different forms of the
symmetry energy/potential predicted by various many-body theories
without changing any property of the symmetric nuclear matter and
the symmetry energy at normal density $\rho_0$. The last two terms
in Eq. (\ref{mdi}) contain the momentum-dependence of the
single-particle potential. The momentum dependence of the symmetry
potential stems from the different interaction strength parameters
$C_{\tau ,\tau ^{\prime }}$ and $C_{\tau ,\tau }$ for a nucleon of
isospin $\tau $ interacting, respectively, with unlike and like
nucleons in the background
fields. More specifically, we use $C_{unlike}=-103.4$ MeV and $%
C_{like}=-11.7 $ MeV. With these parameters, the nucleon isoscalar
potential estimated from $U_{\text{isoscalar}}\approx
(U_{n}+U_{p})/2$ agrees with the prediction of variational
many-body calculations for symmetric nuclear matter
\cite{wiringa,Das03,liba03}, the BHF approach
\cite{bombaci,zuo99,zuo} including three-body forces and the
Dirac-Brueckner-Hartree-Fock (DBHF) calculations \cite{sam05} in
broad ranges of density and momentum. And the corresponding
pressure of symmetric matter is consistent with the experimental
limits \cite{pla08}. The corresponding isovector (symmetry)
potential can be estimated from $U_{sym}\approx
(U_{\text{n}}-U_{\text{p}})/2\delta $.
\begin{figure}[th]
\begin{center}
\includegraphics[width=0.43\textwidth]{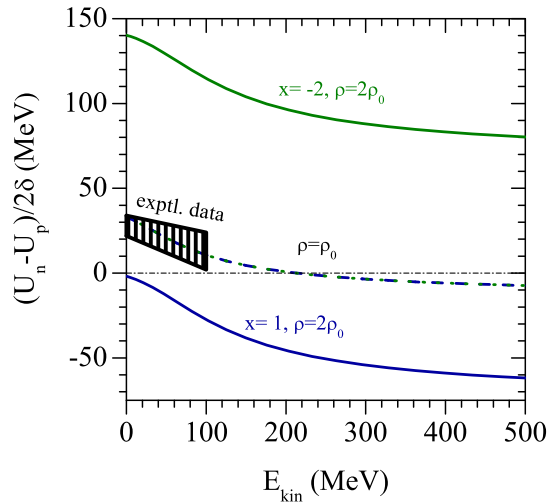}
\end{center}
\caption{The symmetry potential as a function of energy for
different density values in the MDI interaction with $x=1$ and
$x=-2$. The experimental data \cite{data} are also shown.}
\label{psym}
\end{figure}
With different $x$ parameters at normal nuclear matter density
$\rho _{0}$, the symmetry potential, as shown in Fig.~\ref{psym},
agrees very well with the Lane potential extracted from
nucleon-nucleus and (n,p) charge exchange reactions available for
nucleon kinetic energies up to about $100$ MeV \cite{data}. At
supra-saturation densities we can see that the stiff symmetry
energy ($x=-2$, shown in Fig.~\ref{esym}) corresponds to positive
symmetry potential while the soft symmetry energy corresponds to
negative symmetry potential.

According to essentially all microscopic model calculations, see e.g., \cite%
{bombaci,zuo99}, the EOS for isospin asymmetric nuclear matter can
be expressed as
\begin{equation}
E(\rho ,\delta )=E(\rho ,0)+E_{\text{sym}}(\rho )\delta ^{2}+\mathcal{O}%
(\delta ^{4}),
\end{equation}%
where $E(\rho ,0)$ is the energy per nucleon of symmetric nuclear
matter, and $E_{\text{sym}}(\rho )$ is the nuclear symmetry
energy. With the single particle potential Eq.(\ref{mdi}), the
symmetry energy can be written as \cite{Xu09}
\begin{eqnarray}\label{esymmdi}
E_{sym}(\rho, x)&=& \frac{1}{2}\Big(\frac{\partial^{2}E}{\partial\delta^{2}}\Big)_{\delta=0} \nonumber\\
&=&\frac{8 \pi}{9 m h^3 \rho} p^5_f + \frac{\rho}{4%
\rho_0} [-24.59+4Bx/(\sigma +1)] \nonumber\\
&&- \frac{B x}{\sigma + 1} \left(\frac{\rho}{\rho_0}\right)^\sigma
+ \frac{C_{like}}{9 \rho_0 \rho} \left(\frac{4 \pi}{h^3}\right)^2\Lambda^2\nonumber\\
&&\times\left[4 p^4_f - \Lambda^2 p^2_f \ln \frac{4 p^2_f
+ \Lambda^2}{\Lambda^2}\right] \nonumber\\
&&+ \frac{C_{unlike}}{9\rho_0 \rho} \left(\frac{4
\pi}{h^3}\right)^2\Lambda^2\nonumber\\
&&\times\left[4 p^4_f - p^2_f (4 p^2_f + \Lambda^2) \ln
\frac{4p^2_f + \Lambda^2}{\Lambda^2}\right],
\end{eqnarray}
where $p_f=\hbar(3\pi^2\frac{\rho}{2})^{1/3}$ is the Fermi
momentum for symmetric nuclear matter at density $\rho$.
In the mean-field approximation, the symmetry energy
$E_{sym}(\rho, x)$ is actually decided by its potential part
$E_{sym}^{p}(\rho, x)$, $E_{sym}(\rho, x)
=\frac{\hbar}{6m_{n}}(3\pi^{2}\rho/2)^{2/3}+E_{sym}^{p}(\rho, x)$
and the $E_{sym}^{p}(\rho, x)$ is directly related to the symmetry
potential $(U_{\text{n}}-U_{\text{p}})/2\delta$ \cite{liba03}.
Shown in Fig.~\ref{esym} is the density-dependent symmetry energy.
At sub-saturation densities, recent studies have constrained the
symmetry energy around $31.6(\rho/\rho_{0})^{0.69}$ (corresponding
$x=0$) \cite{shetty07,piek08,lit07,todd05}. So in the present
work, we use the symmetry energy/potential corresponding $x=0$ as
our choice at sub-saturation densities. At supra-saturation
densities, because the symmetry energy/potential is very
uncertain, we use the positive ($x=-2$) and negative ($x=1$)
symmetry potential as two extreme cases (corresponding super-stiff
and super-soft symmetry energy at supra-saturation densities,
respectively).

In the IBUU transport model, the initial neutron and proton
density distributions of the projectile and target are obtained by
using the Skyrme-Hartree-Fock theory. The isospin-dependent
in-medium nucleon-nucleon (NN) elastic cross sections from the
scaling model according to nucleon effective masses are used. For
the inelastic cross sections we use the experimental data from
free space NN collisions since the in-medium inelastic NN cross
sections are still very much controversial. The total and
differential cross sections for all other particles are taken
either from experimental data or obtained by using the detailed
balance formula. The isospin dependent phase-space distribution
functions of the particles involved are solved by using the
test-particle method numerically. The isospin-dependence of Pauli
blockings for fermions is also considered, for more details we
refer the reader to Ref. \cite{yong06}.


The nuclear liquid-gas phase transition in dilute asymmetric
nuclear matter is studied recently \cite{liba07}. It argued that
the neutron to proton ratio of the gas phase becomes smaller than
that of the liquid phase for energetic nucleons and the gas phase
is still overall more neutron-rich than that of liquid phase.
Noticing that the recent comparisons between IBUU calculations and
FOPI data favor a rather soft symmetry energy ($x=1$)
(corresponding a negative potential at supra-saturation densities)
at supra-saturation densities \cite{xiao09} and the progress of
the studies of nuclear symmetry energy at sub-saturation densities
\cite{shetty07,piek08,lit07,todd05}, in this work we studied the
$n/p$ of clusters emitted in the central $^{197}$Au+$^{197}$Au
reaction at a beam energy of $400$ MeV/nucleon with the positive
($x=-2$) and negative ($x=1$) symmetry potentials at
supra-saturation densities \emph{but} keeping the symmetry
potential at sub-saturation densities fixed ($x=0$). The maximal
compressed density reached in this reaction is about $2.5\rho_{0}$
\cite{xiao09}.
\begin{figure}[th]
\begin{center}
\includegraphics[width=0.5\textwidth]{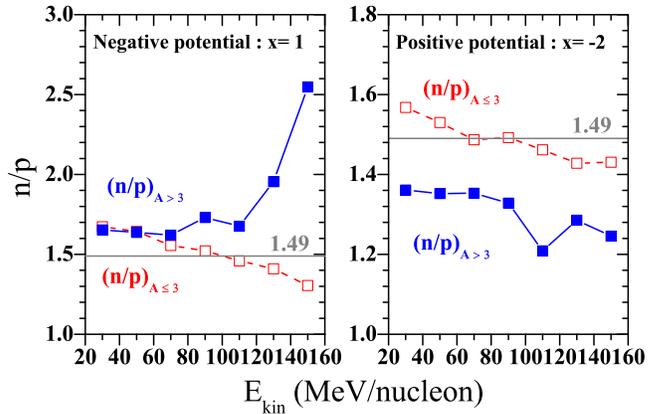}
\end{center}
\caption{Neutron to proton ratio $n/p$ as a function of nucleonic
kinetic energy of clusters with $A\leq3$, $A>3$ from the central
$^{197}$Au+$^{197}$Au reaction at a beam energy of $400$
MeV/nucleon with the positive ($x=-2$) and negative ($x=1$)
symmetry potentials at supra-saturation densities but keeping the
symmetry potential at sub-saturation densities fixed ($x=0$). The
number $1.49$ denotes the reaction system's $n/p$.} \label{phase}
\end{figure}
Shown in Fig.~\ref{phase} is $n/p$ of clusters with $A\leq3$,
$A>3$ as a function of nucleonic kinetic energy. For the negative
symmetry potential, neutrons trend to being attracted by the
symmetry potential and protons trend to being repelled during
isospin fractionation. Thus $n/p$ of weighter clusters ($A>3$) are
larger than that of the lighter clusters ($A\leq3$). Whereas for
positive symmetry potential, neutrons trend to being repelled by
the symmetry potential and protons trend to being attracted during
isospin fractionation. Therefore $n/p$ of weighter clusters
($A>3$) are smaller than that of the lighter clusters ($A\leq3$).
From Fig.~\ref{phase}, we can also see that $n/p$ of weighter
clusters ($A>3$) with negative symmetry potential is overall
larger than reaction system's $n/p$ while with the positive
symmetry potential $n/p$ of weighter clusters ($A>3$) is expected
overall smaller than reaction system's $n/p$. In Fig.~\ref{phase},
$(n/p)_{A\leq3}$ and $(n/p)_{A>3}$ are the neutron to proton
ratios of nucleons with local densities smaller ($A\leq3$) and
larger ($A>3$) than $\rho_{c}=1/8\rho_{0}$, respectively. Changing
$\rho_{c}$ from $1/5\rho_{0}$ to $1/10\rho_{0}$, our quantitative
results only shift about 3\%. The integral
$(n/p)_{A\leq3}$/$(n/p)_{A>3}$ with nucleonic kinetic energy
$E_{kin}/nucleon\geq$ 20 MeV (nucleons with kinetic energy
$E_{kin}/nucleon< 20$ MeV are mainly from cluster decays
\cite{fami06}) is $1.44/1.65$ for the negative symmetry potential
and $1.48/1.35$ for the positive symmetry potential.
\begin{table}[th]
\caption{The clusterisation method and the associated parameters.}
\label{notef3}%
\begin{tabular}{|c|c|c|c|}
  \hline
  set & $n/p$ & $\rho_{c}=1/8\rho_{0}$ & $P_{0}=263$ MeV/c \\
  \hline
  Negative potential & $A\leq3$ & 1.44 & 1.44 \\
  (x= 1) & $A>3$ & 1.65 & 1.55 \\
  \hline
  Positive potential & $A\leq3$ & 1.48 & 1.49 \\
  (x= -2) & $A>3$ & 1.35 & 1.42 \\
  \hline
\end{tabular}
\end{table}
In the standard implementation of cluster recognition after the
BUU dynamics \cite{yon09}, a physical fragment is formed as a
cluster of nucleons with relative momenta smaller than $P_{0}=263$
MeV/c (Fermi-Momentum of normal nuclear matter) and relative
distances smaller than $R_{0}=3$ fm (deduced by the uncertainty
relationship of quantum mechanics). The integral
$(n/p)_{A\leq3}$/$(n/p)_{A>3}$ with nucleonic kinetic energy
$E_{kin}/nucleon\geq 20$ MeV is $1.44/1.55$ for the negative
symmetry potential and $1.49/1.42$ for the positive symmetry
potential and our results are not sensitive to the phase-space
coalescence parameter settings (when changing $P_{0}$ from 263
MeV/c to $P^{F}_{Au}\approx 258$ MeV/c, the results shift not more
than 0.3\%). In order to show more clearly the above results, We
present Table \ref{notef3}. It is seen that the clusterisation
method based on phase-space coalescence and the method based on a
density cut-off give the same $(n/p)_{A\leq3}$, but different
$(n/p)_{A>3}$. Specifically, the symmetry energy dependence of the
isospin content of large clusters is less pronounced with the
coalescence method. This can be qualitatively understood. Indeed
the nuclear Fermi momentum of asymmetric nuclear matter is
$p_{F}^{p}=(1-\delta)^{1/3}p_{0},~~p_{F}^{n}=(1+\delta)^{1/3}p_{0}$.
Where $p_{F}^{p}, p_{F}^{n}$ are proton and neutron's
Fermi-Momenta, respectively \cite{zuo99}. At densities close to
saturation neutrons will then have a momentum that can exceed the
coalescence parameter $P_{0}$ which corresponds to symmetric
nuclear matter at saturation. Therefore in average the coalescence
criterium will be harder to fulfill for neutrons than for protons
at the same density. This can explain why, for the negative
symmetry potential, $(n/p)_{A>3}$ is lower with the coalescence
method than with the density cut-off method. The opposite effect
is observed with the positive symmetry potential, which leads to
more neutron-rich dense matter. From the above we can see that
although the two methods (the clusterisation method based on
phase-space coalescence and the method based on a density cut-off)
give different quantitative results, they give the same
qualitative results, i.e., $(n/p)_{A\leq3}$ is larger/smaller than
$(n/p)_{A>3}$ for the positive/negative symmetry potential at
supra-saturation densities.

The overall $(n/p)_{A\leq3}$ is smaller than $(n/p)_{A>3}$ with
nucleonic kinetic energy $E_{kin}/nucleon\geq 20$ MeV for the
negative symmetry potential at supra-saturation densities after
isospin fractionation is different from the knowledge in the
literatures \cite{das,Bar05,chomaz,li00,xu,liba07} for lower
energy density nuclear isospin fractionation where mainly the
symmetry energy/potential at sub-saturation densities works. In
case experimentally in the central $^{197}$Au+$^{197}$Au reaction
at an incident beam energy of $400$ MeV/nucleon $(n/p)_{A\leq3}$
is overall smaller than $(n/p)_{A>3}$ with nucleonic kinetic
energy $E_{kin}/nucleon\geq 20$ MeV, then a negative symmetry
potential at supra-saturation densities is obtained. Otherwise the
negative symmetry potential at supra-saturation densities is ruled
out. This test in fact does not dependent on our quantitative
results of $(n/p)_{A\leq3}$ or $(n/p)_{A>3}$. This can be a probe
of the negative symmetry potential at supra-saturation densities.
It should be mentioned that different isospin dependent in-medium
nucleon-nucleon cross sections may affect our present results
\cite{lich05,liq09}, such studies are planned.


In summary, based on an isospin dependent transport model IBUU,
$n/p$ of the clusters as a test of negative symmetry potential at
supra-saturation densities is studied. We found that for the
positive symmetry potential at supra-saturation densities the
neutron to proton ratio of lighter clusters with mass number
$A\leq3$ is larger than that of the weighter clusters with mass
number $A>3$, but for the negative symmetry potential at
supra-saturation densities the $(n/p)_{A\leq3}$ is \emph{smaller}
than that of the $(n/p)_{A>3}$. This may be considered as a probe
of the negative symmetry potential at supra-saturation densities.

The author thanks the referees for the comments on the manuscript.
This work is supported in part by the National Natural Science
Foundation of China under grants 10740420550, 10875151.


\begin{thebibliography}{00}

\bibitem{Bro00}B.A. Brown, Phys. Rev. Lett. {\bf 85}, 5296 (2000).

\bibitem{Bar05}V. Baran et al., Phys. Rep. {\bf 410}, 335 (2005).

\bibitem{LCK08}B.A. Li, L.W. Chen and C.M. Ko, Phys. Rep. {\bf 464}, 113 (2008).

\bibitem{Sum94}K. Sumiyoshi and H. Toki, Astrophys. J. {\bf 422}, 700 (1994).

\bibitem{Lat04}J.M. Lattimer, M. Prakash, Science {\bf 304}, 536 (2004).

\bibitem{Ste05a}A.W. Steiner et al., Phys. Rep. {\bf 411}, 325 (2005).

\bibitem{piek08}J. Piekarewicz, Phys. Rev. {\bf C76}, 064310
(2007); A. Schwek and C.J. Pethick, Phys. Rev. Lett. {\bf 95},
160401 (2005).

\bibitem{horow01}C.J. Horowitz, J. Piekarewicz, Phys. Rev. Lett. {\bf
86}, 5647 (2001).

\bibitem{xiao09}Z.G. Xiao et al., Phys. Rev. Lett. {\bf 102}, 062502 (2009).

\bibitem{yong06}G.C. Yong et al., Phys. Rev. {\bf C73}, 034603 (2006).

\bibitem{LiBA02}B.A. Li, Phys. Rev. Lett. {\bf 88}, 192701 (2002).

\bibitem{Kut94}M. Kutschera, Phys. Lett. {\bf B340}, 1 (1994).

\bibitem{Kub99}S. Kubis and M. Kutschera, Acta Phys. Pol. {\bf B30},
2747 (1999); Nucl. Phys. {\bf A720}, 189 (2003).

\bibitem{tsang09}M.B. Tsang et al., Phys. Rev. Lett. {\bf 102}, 122701 (2009).

\bibitem{shetty07}D.V. Shetty et al., Phys. Rev. {\bf C76}, 024606 (2007)
and references therein.

\bibitem{fami06}M.A. Famiano et al., Phys. Rev. Lett. {\bf 97}, 052701 (2006).

\bibitem{tsang04}M.B. Tsang et al., Phys. Rev. Lett. {\bf 92}, 062701 (2004).

\bibitem{chen05}L.W. Chen et al., Phys. Rev. Lett. {\bf 94}, 032701 (2005).

\bibitem{liba03}B.A. Li et al., Nucl. Phys. {\bf A735}, 563 (2004).

\bibitem{fuchs}E.N.E. van Dalen, C. Fuchs, A. Faessler, Nucl. Phys. {\bf A744}, 227 (2004).

\bibitem{zuo}W. Zuo et al., Phys. Rev. {\bf C72}, 014005 (2005).

\bibitem{beh05}B. Behera et al., Nucl. Phys. {\bf A753}, 367 (2005).

\bibitem{liqf05}Qingfeng Li et al., Phys. Rev. {\bf C72}, 034613 (2005).

\bibitem{liz06}Z.H. Li et al., Phys. Rev. {\bf C74}, 044613 (2006).

\bibitem{chomaz}Ph. Chomaz et al., Phys. Rep. {\bf 389}, 263 (2004).

\bibitem{das}C.B. Das et al., Phys. Rep. {\bf 406}, 1 (2005).

\bibitem{li00}B.A. Li, Phys. Rev. Lett. {\bf 85}, 4221 (2000).

\bibitem{xu}H.S. Xu et al., Phys. Rev. Lett. {\bf 85}, 716 (2000).

\bibitem{Das03}C. B. Das et al., Phys. Rev. {\bf C67}, 034611 (2003).

\bibitem{bombaci}I. Bombaci and U. Lombardo, Phys. Rev. {\bf C44}, 1892 (1991).

\bibitem{k0data}D.H. Youngblood et al., Phys. Rev. Lett. {\bf 82}, 691 (1999).

\bibitem{pie04}J. Piekarewicz, Phys. Rev. {\bf C69}, 041301 (2004).

\bibitem{colo04}G. Colo et al., Phys. Rev. {\bf C70}, 024307 (2004).

\bibitem{wiringa}R.B. Wiringa, Phys. Rev. {\bf C38}, 2967 (1988).

\bibitem{zuo99}W. Zuo, I. Bombaci and U. Lombardo, Phys. Rev. {\bf C60}, 024605 (1999).

\bibitem{sam05}F. Sammarruca, W. Barredo and P. Krastev, Phys. Rev.
{\bf C71}, 064306 (2005).

\bibitem{pla08}P.G. Krastev, B.A. Li, A. Worley, Phys. Lett. {\bf B668}, 1
(2008); P. Danielewicz, R. Lacey, W.G. Lynch, Science {\bf 298},
1592 (2002).

\bibitem{data}P.E. Hodgson, The Nucleon Optical Model, pages
613-651, (World Scientific, Singapore, 1994).

\bibitem{Xu09}J. Xu, L.W. Chen, B.A. Li and H.R. Ma, Astrophys. J. 697, 1549
(2009).

\bibitem{lit07}T. Li et al., Phys. Rev. Lett. {\bf 99}, 162503 (2007).

\bibitem{todd05}B.G. Todd-Rutel, J. Piekarewicz, Phys. Rev. Lett. {\bf 95}, 122501 (2005).

\bibitem{liba07}B.A. Li et al., Phys. Rev. {\bf C76}, 051601 (2007).

\bibitem{yon09}G.C. Yong et al., Phys. Rev. {\bf C80}, 044608
(2009); H. Kruse et al., Phys. Rev. {\bf C31}, 1770 (1985).

\bibitem{lich05}B.A. Li and L.W. Chen, Phys. Rev. {\bf C72}, 064611 (2005).

\bibitem{liq09}Qingfeng Li, Caiwan Shen, M. Di Toro,  arXiv: 0908.2825 (2009).


\end{thebibliography}
\end{document}